\documentclass[prl,aps,twocolumn,groupedaddress,floats,showpacs,final,superscriptaddress]{revtex4-1}
\usepackage[latin3]{inputenc}
\usepackage[makeroom]{cancel}
\usepackage{graphicx}
\usepackage{amsmath}
\usepackage{amsfonts}
\usepackage{amssymb}
\usepackage{color}
\usepackage[left=2cm,right=2cm,top=2cm,bottom=2cm]{geometry}

\usepackage{graphicx} 
\usepackage{dcolumn}
\usepackage{bm}
\usepackage{simplewick}
\usepackage{array}
\usepackage{appendix}

\RequirePackage[
   hyperindex,colorlinks,bookmarksnumbered,
   plainpages=true,pdfstartview=FitH]{hyperref}
\hypersetup{linkcolor=blue,urlcolor=blue,citecolor=blue}
\usepackage{hyperref}

\definecolor{purple}{rgb}{0.5,0,0.6}

\usepackage{ulem}
\renewcommand{\emph}[1]{\textit{#1}}
\definecolor{darkblue}{rgb}{0,0,0.5}
\definecolor{darkgreen}{rgb}{0,0.5,0}
\definecolor{darkred}{rgb}{.7,0,0}
\definecolor{purple}{rgb}{0.5,0,0.6}
\definecolor{orange}{rgb}{1,0.5,0}
\definecolor{grey}{rgb}{.6,.6,.6}
\definecolor{lightpink}{rgb}{1,0.7,0.75}
\definecolor{pink}{rgb}{1,0.4,0.58}
\definecolor{deeppink}{rgb}{1,0.08,0.58}

\newcommand{\DK}[1]{{\color{black}{#1}}} 

\renewcommand{\emph}[1]{\textit{#1}}

\begin{document}


\title{Effects of strong electron interactions and resonance scattering
on power output of nano-devices}



\author{D. B. Karki}
\affiliation{International School for Advanced Studies (SISSA), Via Bonomea 265, 34136 Trieste, Italy}
\affiliation{The  Abdus  Salam  International  Centre  for  Theoretical  Physics  (ICTP)
Strada  Costiera 11, I-34151  Trieste,  Italy}
\author{Mikhail N. Kiselev}
\affiliation{The  Abdus  Salam  International  Centre  for  Theoretical  Physics  (ICTP)
Strada  Costiera 11, I-34151  Trieste,  Italy}



\begin{abstract}
\DK{We develop a Fermi-liquid based approach to investigate the power output of nano devices in the presence of strong interactions and resonance scattering. The developed scheme is then employed to study the power output of a SU($N$) Kondo impurity at the strong-coupling regime. The interplay between Kondo resonance and the filling-factors in the SU($N$) quantum systems is found to be a key to enhance output power. Such enhancement results an output power corresponding to $50\%$ of the quantum upper bound. We demonstrate that given a proper tuning of the electron occupancy, the investigated power grows linearly with degeneracy of Kondo state ($N$). This relation can hence be exploited to obtain output power that is larger than the one in existing non interacting setups.}
\end{abstract}

\maketitle
\textit{Introduction}$-$ Greatly enhanced thermoelectric response of nano scale systems over conventional bulk materials has revived further the field of thermoelectricity~\cite{ld00, ld0, ld1}. Rapid development of nanotechnology have fueled several exciting thermoelectric experiments on nano materials and their theoretical formulation~\cite{casti} to fulfill the urgent demand of energy harvesters for quantum technologies. The charge quantization in quantum devices~\cite{leo} furnishes a controllable comprehension of underlying transport processes. Consequently, spectacular thermoelectric measurement of prototypical nano scale systems such as quantum dots (QDs), carbon nano-tubes (CNTs), quantum point contacts (QPCs), etc has been reported over the past years~\cite{ld00, Blanter}. Natural consequences of being the systems size at the nano scale imply the ubiquitousness of electron interactions. Equivalently, strong Coulomb interaction is at the cornerstone of nano devices. The resonance scattering often combined with strong electron interaction which drives the system to posses very peculiar functionality~\cite{leo3, leo2}. Therefore unified description of resonance scattering and strong electrons interaction at the nano scale have remain challenging task for modern quantum technologies. 

In the past years several perseverance has been devoted for the consistent description of thermoelectricity in QD based heat engines~\cite{casti}. Efficiency and power production of a heat engine are the two connected fundamental ingredients of thermoelectric production~\cite{cal,esp0,espl, esp}. Reversible engine, though Carnot efficient, are not of any practical applications since they
do not produce finite power~\cite{cal}. Consequently the search for quantum thermoelectric devices with maximum attainable output power maintaining good efficiency has remain one of the active and demanding field of research in mesoscopic physics~\cite{esp, casti}. Even though unveiling the universal upper bound of output power \DK{of a generic} nano devices looks like a serious challenge, certain attempts of this facet has been reported recently~\cite{whitney1, whitney2, whitney3}. These fundamental discoveries by \DK{Robert Whitney}, for the first time, have shown that the quantum mechanics sets an upper bound on the power output of the non-interacting systems.
Based on the non-linear scattering theory, Whitney predicted the maximum power output of two-terminal nano devices in the form~\cite{whitney1} ${\rm P}^0_{\rm max}\leq \mathcal{K}/h\times A_0 \pi^2 \Delta T^2,  A_0\simeq 0.0321$, $\mathcal{K}$ is the number of transverse modes participating in the transport, $h$ is the Planck's constant and $\Delta T$ is the applied temperature gradient. \DK{Note that, the recent experiment~\cite{kancho1} has reported the output power only about $50\%$ of quantum bound in one dimensional nanowire. Likewise, $75\%$ of this quantum bound has been suggested in the recent work~\cite{notes}.}

\begin{figure}[t]
\includegraphics[scale=0.34]{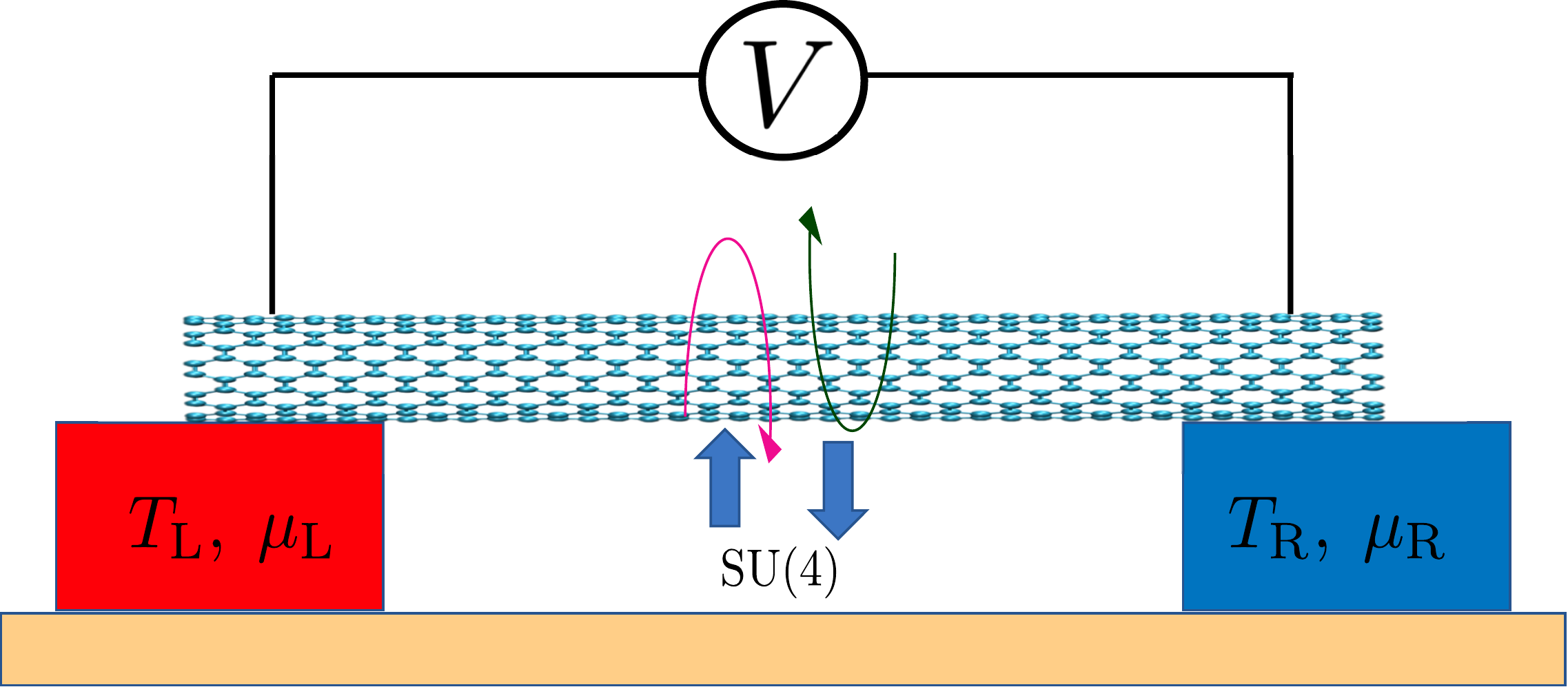}
\caption{An example of SU($N{=}4$) Kondo correlated heat engine where a CNT is connected between two fermionic reservoirs, the hot (red) and the cold (blue). Voltage bias $\Delta V$ and temperature gradient $\Delta T{=}T_{\rm L}{-}T_{\rm R}$ is applied across the CNT quantum dot. The doubly degenerate orbital degree of freedom in CNT combines with true spin degeneracy to form a Kondo effect described by the SU(4) symmetry group.}\label{setup}
\vspace*{-5mm}
\end{figure}

In addition to greatly contributing towards the better understanding of nano scale thermoelectricity, Whitney's finding has open diverse valid avenues for future research, both theoretical and experimental. The examinations of how universal is this bound ${\rm P}^0_{\rm max}$ in the presence of strong electron interactions,\DK{ such as with the Kondo physics,} are of earnest interest~\cite{whitney2}. \DK{In addition, since the quantum bound of output power is linearly growing function of $\mathcal{K}$, the efficiency at given power output can be significantly increased with increasing $\mathcal{K}$. However, as pointed out by Whitney in his original work~\cite{whitney1}, most quantum thermoelectric devices are often limited to the setup with $\mathcal{K}{\sim}1$. The experimental search of achieving $\mathcal{K}{\gg}1$ to increase the total power output of a nano device has remains one of the active field of research~\cite{casti}. 

The strong electron interactions at the nano scale often lead to the paradigmatic Kondo screening phenomena~\cite{kondo} which plays a central role of enhancing nano scale thermoelectric production~\cite{kiselev, if3, if1, if2, if4, heiner, paulo}. The low temperature Kondo regime emerges from the complete screening of spin of the localize impurity forming a strongly correlated Fermi-liquid (FL) ground state~\cite{Nozieres}. The conventional SU(2) Kondo effects, being protected by particle-hole (PH) symmetry (half-filling), results in vanishingly small power output~\cite{su2_1, su2_2, su2_3, kiselev,costi1, dee1}. Nonetheless, the SU($N{>}2$) Kondo effects offers a nontrivial occupation away from the half-filling opens the possibility of achieving a giant power production. The strong interplay of Kondo resonance and filling factor causes the SU($N{>}2$) Kondo effects to posses dramatic transport features at low temperature as revealed by recent experimental~\cite{jh, sasa0, sasa, if2, if4, keller, st} and theoretical~\cite{hur,su4_21, su4_22, su4_23, su4_11, aash, rok1, kita, su12, nashida1, nashida2, su6, yemkali} perseverances.}

\DK{A generic nano device in SU($N$) Kondo regime, therefore, allows us: i) to utilize the enhanced electronic features of Kondo effect with the possibility of detuning from the PH symmetric point and ii) to achieve $\mathcal{K}{\gg} 1$ in Whitney's upper bound of power output to approach towards the Carnot efficiency~\cite{whitney1}. Note that, the degeneracy of Kondo ground state $N$ with SU($N$) quantum impurity plays an equivalent role of $\mathcal{K}$ in the Whitney's formula. In non-interacting systems, the consideration of spin degeneracy can just double the charge current resulting in $\mathcal{K}{=}2$. Therefore, utilizing SU($N$) Kondo resonance resulting from the strong electron interactions would be a highly beneficial way of giant power production with a nano device. This argument of using SU($N$) Kondo resonance to achieve $\mathcal{K}{\gg}1$ is very different than Whitney's suggestion of using many properly engineered quantum systems in parallel~\cite{whitney1, whitney2}.

In this Letter, we develop a theoretical framework based on the local FL approach to investigate the output power of a generic nano device in the strong-coupling regime of a SU($N$) Kondo impurity. The developed formalism is employed to reveal the influences of resonance scattering and strong interaction on the power output of a nano devices. The investigated linear response upper bound of power production in SU($N$) Kondo regime is analyzed in relation to the corresponding non-interacting upper bound established by Whitney~\cite{whitney1}. To obtain a giant output power that scales linearly with $N$, we utilize the Kondo effect being at the same time away from the half-filling. We made close connection between our finding and experimentally studied SU(4) Kondo effects in CNTs and double QDs.}

\textit{Theoretical formulation}$-$ We consider a SU($N$) Kondo impurity tunnel coupled to two conducting reservoirs as shown in Fig.~\ref{setup}. \DK{The reservoirs are also assumed to be described by the same symmetry group SU($N$)~\footnote{For instance, the SU(4) symmetry is achieved by depositing CNT on the top of both electronic reservoirs.}.} In addition, the left and right reservoirs are in equilibrium, separately, at temperatures $T_{\gamma}$ ($\gamma{=}\rm {L, R}$) and chemical potentials $\mu_{\gamma}$ respectively. Note that throughout the calculations we use
the system of atomic unit, $\hbar{=}k_B{=}e{=}1$ unless explicitly
written for special propose. As far as the condition $T_{\rm L}{\neq} T_{\rm R}$ and $\mu_{\rm L}{\neq} \mu_{\rm R}$ is satisfied, heat current ($I_{\rm h}$) and charge current ($I_{\rm c}$) start to flow from one reservoir to another via the Kondo impurity. To be more explicit, we consider the voltage bias and temperature gradient across the impurity in such a way that $\mu_{\rm L}{-}\mu_{\rm R}{=}\Delta V$ and $T_{\rm L}{-}T_{\rm R}{=}\Delta T$. We choose the right reservoir to define the reference temperature, $T_{\rm ref}{\equiv}T_{\rm R}{=}T$. Then the charge and the heat currents in the linear response theory are connected by the Onsagar relations~\cite{sagar1, sagar2}, 
\begin{equation}\label{bd0}
 \left(%
\begin{array}{c}
  I_{\rm c} \\
  I_{\rm h} \\
\end{array}%
\right)= \left(%
\begin{array}{cc}
  {\rm L}_{11} & {\rm L}_{12} \\
  {\rm L}_{21} & {\rm L}_{22} \\
\end{array}%
\right)\left(%
\begin{array}{c}
  \Delta V \\
  \Delta T \\
\end{array}%
\right).
\end{equation}
The Onsagar transport coefficients ${\rm L_{\rm ij}}$ in Eq.~\eqref{bd0} provide all the thermoelectric measurements of interests in linear response regime~\cite{costi1}. As we anticipated earlier, the low energy transport via fully screened SU($N$) Kondo impurity is completely described by a local FL theory~\cite{Nozieres}. Therefore the coefficients ${\rm L_{\rm ij}}$ are characterized by the single particle T-matrix $T_{\sigma}(\varepsilon)$ of FL quasi-particles~\cite{GP_Review_2005, costi1}. Such connection is governed by defining the transport integrals ${\cal I}_{\rm n}(T)$ (${\rm n}{=}0, 1\; \text{and}\;2$) in terms of the imaginary part of the T-matrix~\citep{GP_Review_2005},
\begin{equation}\label{bd2}
{\cal I}_n(T){=}\sum_{\rm \sigma}\int^{\infty}_{-\infty} \frac{d\varepsilon}{2\pi}\; \varepsilon^n\left[-\frac{\partial f(\varepsilon)}{\partial \varepsilon}\right]\;
{\rm Im}\left[-\pi \nu T_\sigma(\varepsilon)\right].
\end{equation}
Here $f(\varepsilon){=}\left[1+\exp\left(\varepsilon/T\right)\right]^{-1}$ is the equilibrium Fermi-distribution function of the reference reservoir. The orbital index is represented by  the symbol $\sigma$ which takes all possible values starting from 1 to $N$. Density of states per species for the one-dimensional channel $\nu$ and the Fermi-velocity $v_F$ are related as $\nu=1/2\pi v_F$. Then the transport coefficients characterizing the charge current are expressed in terms of the transport integrals, namely, ${\rm L_{11}}=\mathcal{I}_0$ and ${\rm L_{12}}=-\mathcal{I}_1/T$~\cite{kim}. This reduces our task just to find the expression of the single particle T-matrix of Eq.~\eqref{bd2} using a local FL-paradigm. To proceed with the calculation of T-matrix, we considered that ${\cal G}^0_{k\sigma}(\varepsilon)$ and ${\cal G}_{k\sigma}(\varepsilon)$ represents the bare and full Green's functions (GFs) of FL quasi-particles. In addition, we consider the
$k$-independence of the T-matrix which is valid for the local interactions considered in this work. Then the diagonal part of a single-particle T-matrix $T_\sigma(\varepsilon)$ is defined by the equation \cite{GP_Review_2005},
\begin{equation}\label{bd4}
\mathcal{G}_{k\sigma, k'\sigma'}(\varepsilon)=\delta_{\sigma\sigma'}\mathcal{G}^0_{k\sigma}(\varepsilon)+\mathcal{G}^0_{k\sigma}(\varepsilon){T}_{\sigma\sigma'}(\varepsilon)\mathcal{G}^0_{k'\sigma'}(\varepsilon).
\end{equation}
Here $\delta_{\sigma\sigma'}$ is the Kronecker delta function. Fourier transforming Eq.~\eqref{bd4} into the position space and applying the particle conservation constraint along with the fact that low-energy electron cannot flip spin due to the scattering events, we get the elastic part of T-matrix as
\begin{align}\label{bd6}
-\pi\nu{T}^{\rm el}_{\sigma}(\varepsilon)=\frac{1}{2i}\left[e^{2i\delta^{\rm el}_{\sigma}(\varepsilon)}-1\right].
\end{align}
In the Eq.~\eqref{bd6} we expressed the elastic part of the scattering matrix $\mathcal{S}^{\rm el}_{\sigma}(\varepsilon)$ in terms of the scattering phase shift $\delta^{\rm el}_{\sigma}(\varepsilon)$ such that $\mathcal{S}^{\rm el}_{\sigma}(\varepsilon)\equiv e^{2i\delta^{\rm el}_{\sigma}(\varepsilon)}$. The interaction in the FL must be treated perturbatively with small parameters $(\Delta V, \Delta T, T)/T^{\rm SU(N)}_K$, for $T^{\rm SU(N)}_K$ being the Kondo temperature of SU($N$) quantum impurity. However, the Hartree contribution of self-energy can be included in the elastic phase shift~\cite{mora1}. For the SU($N$) Kondo impurity the equilibrium phase shift accounting for the scatting effects and Hartree contribution to the self energy is written in terms of the Nozieres FL parameters~\cite{Nozieres, mora1} 
\begin{equation}\label{bd7}
\delta^{\rm el}_{\sigma} (\varepsilon)=\delta_0+\alpha_1\varepsilon+\alpha_2\left[\varepsilon^2- \frac{(\pi T)^2}{3}\right].
\end{equation}
This general expression for the elastic phase shift is applicable to the strong-coupling regime of SU($N$) Kondo impurity with upto $m{=}N{-}1$ electrons. The phase shift corresponds to the perfect transmission is given by $\delta_0=m\pi/N$~\cite{mora1}. The first and second generations of FL coefficients, $\alpha_1$ and $\alpha_2$ respectively, are related to the Kondo temperature of the system~\cite{dee1}. For the sake of simplicity, we defined the Kondo temperature such that $T^{\rm SU(N)}_K{=}1/\alpha_1$, the $N$-dependence in the FL parameters is implicit. The exact relation between $\alpha_1$ and $\alpha_2$ is given by the Bethe-Ansatz solution  \cite{mora2}:
\begin{equation}\label{bd8}
\frac{\alpha_2}{\alpha^2_1}=\frac{N-2}{N-1}\frac{\Gamma(1/N)\tan(\pi/N)}{\sqrt{\pi}\Gamma\left(\frac{1}{2}+\frac{1}{N}\right)}\cot\left[\frac{m\pi}{N}\right].
\end{equation}
Where $\Gamma(x)$ is the Euler's gamma-function. Note that for the half-filled systems, $m{=}N/2$ (with even $N$), the second generation of the FL-coefficients gets nullified. The T-matrix accounting for the inelastic effects (leaving aside the corresponding Hartree contributions) associated with the quasi-particle interaction in FL theory is given by~\cite{Nozieres, aff}
\begin{equation}\label{bd10a}
T^{\rm in}_{\sigma}(\varepsilon)=\frac{N-1}{2i\pi\nu}\left[\varepsilon^2+(\pi T)^2\right]\phi^2.
\end{equation}
Here $\phi$ is the FL-coefficient representing the interaction effects originated from the four-fermions interaction, the interactions beyond four fermions is neglected for the description of low-energy transport processes. It has been proved that the coefficient $\phi$ is related with $\alpha_1$ such that $\alpha_1{=}(N-1)\phi$~\cite{mora2}. In addition, the inelastic part of the T-matrix Eq.~\eqref{bd10a} is an even function of energy, that is the inelastic transmission function is symmetric with respect to the energy. Such a perfect symmetry tends to nullify the thermoelectric response as will be clear in the following section. Therefore in the linear response level of calculations the thermo-conductance is solely governed by the scattering effects associated with the FL quasi-particles plus the Hartree contribution to the self energy. The T-matrix accounting for the scattering and interactions in the FL is expressed as~\citep{GP_Review_2005},
\begin{equation}\label{bd10b}
T^{\rm tot}_{\sigma}(\varepsilon)\equiv T_{\sigma}(\varepsilon)=T^{\rm el}_{\sigma}(\varepsilon)+e^{2i\delta_0} T^{\rm in}_{\sigma}(\varepsilon).
\end{equation}
Note that the expression of ${-}\pi\nu{\rm Im}T_{\sigma}(\varepsilon)$ in Eq.~\eqref{bd2} contains the cosine factor $\cos2\delta_0$ coupled with the inelastic part Eq.~\eqref{bd10a}. Interestingly, the factor $\cos2\delta_0$ dramatically modifies some transport behaviors. For the quarter filled SU($N$) impurity such that, $m/N=1/4\;\text{or}\;3/4$, the imaginary part of second term in Eq.~\eqref{bd10b} vanishes. Such systems are merely described by the phase shift expression Eq.~\eqref{bd7}. This ideal situation corresponds to the Kondo effects in CNT (see Fig~\ref{setup}), where the SU($4$) Kondo effect comes into play with $m{=}1, 2\;\text{or}\;3$. While the $m=2$, SU($4$) systems have poor thermoelectric performance due to the emergent PH symmetry, the systems of SU($4$) impurity beyond half-filled regime are the ideal test-bed for the study of transport behavior. Use of T-matrix expression given in Eq.~\eqref{bd10b} into the transport integrals Eq.~\eqref{bd2} followed by the Taylor series expansion upto second order in energy yields,
\begin{align}\label{bd11}
&{\cal I}_n(T)=\frac{G^{\rm SU(N)}_0}{4T}\int^{\infty}_{-\infty}d\varepsilon \frac{\varepsilon^n}{\cosh^2\left(\varepsilon/2T\right)}\;\Big[\sin^2\delta_0\nonumber\\
&{+}\frac{(\pi T)^2}{2N-2}\left(\alpha^2_1\cos2\delta_0-\frac{2N-2}{3}\;\alpha_2\sin2\delta_0\right)\nonumber\\
&{+}\alpha_1\sin2\delta_0\varepsilon{+}\left(\frac{2N{-}1}{2N{-}2}\;\alpha^2_1\cos2\delta_0{+}\alpha_2\sin2\delta_0\right)\varepsilon^2\Big].
\end{align}
Here $G^{\rm SU(N)}_0{=}N/2\pi$ is the unitary conductance of SU($N$) system.
\begin{figure}
\includegraphics[scale=0.6]{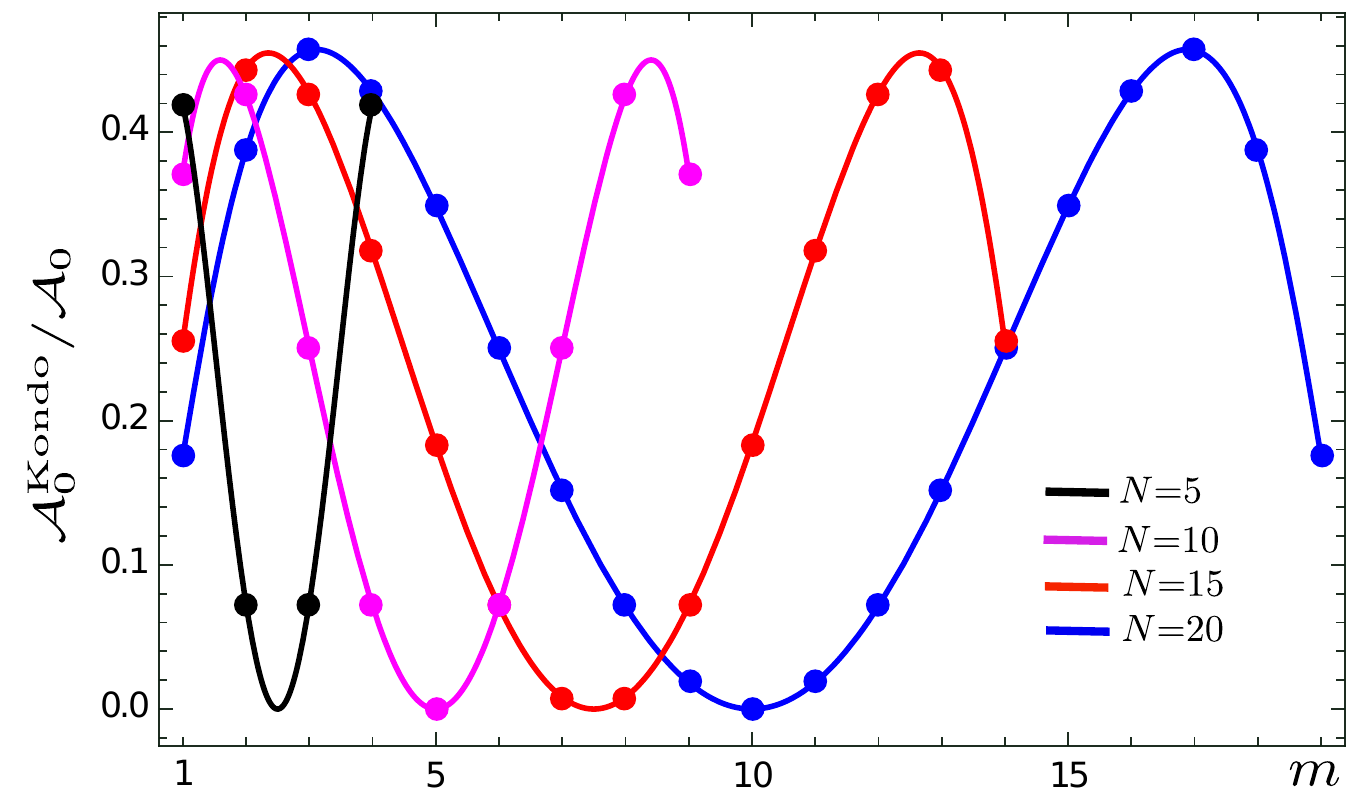}\;\;\;
\includegraphics[scale=0.6]{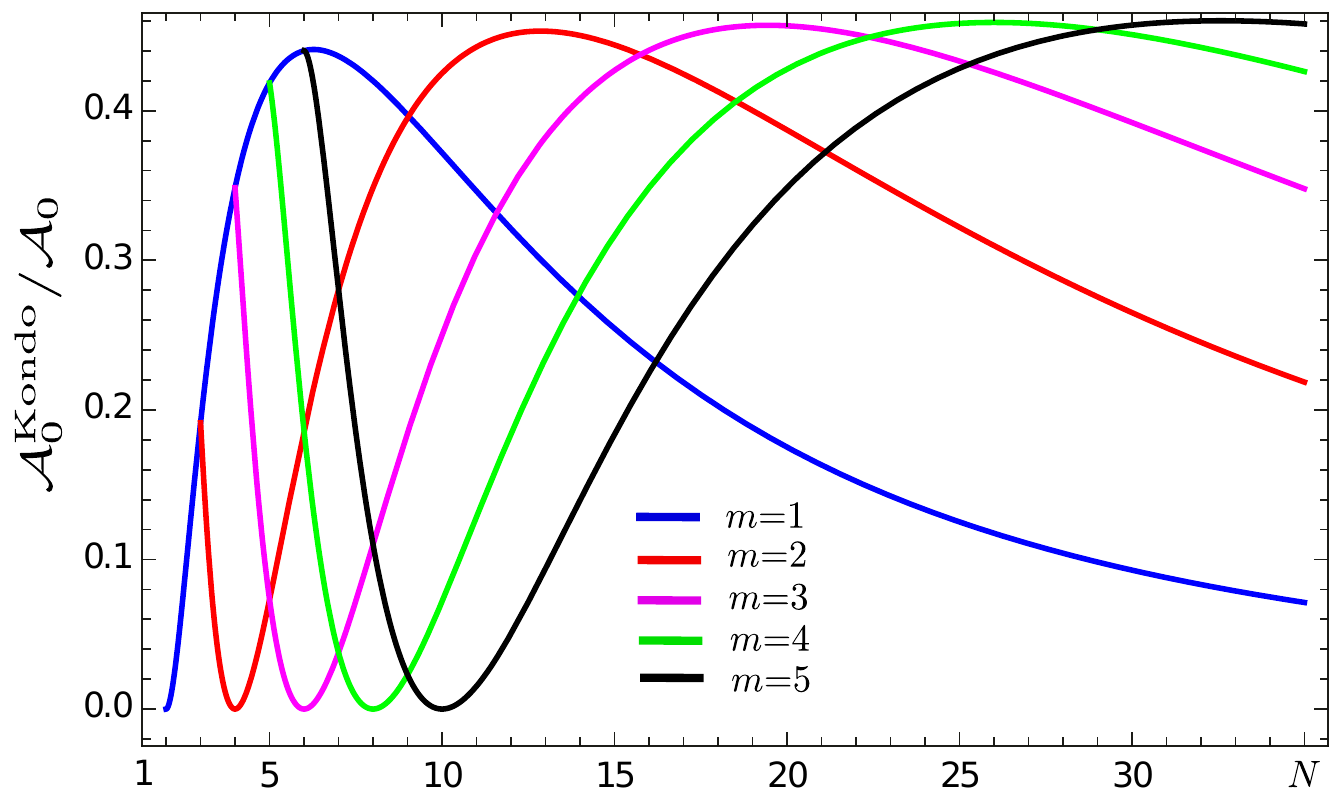}
\caption{\DK{Upper panel: The output power of SU($N$) Kondo impurity per degeneracy in the unit of corresponding quantum upper bound as a function of occupancy $m$ for fixed $N$. Lower panel: The decay of output power of SU($N$) Kondo impurity with $N$ for given $m$. For both plots the reference temperature has been fixed to $T{=}T^{\rm SU(N)}_K/7$.}}\label{exceeding}
\vspace*{-5mm}
\end{figure}

\textit{Results and discussion}$-$ Though all the fundamental measure of thermoelectricity can be extracted from Eq.~\eqref{bd11}, here we shall focus only on the power production. Any thermoelectric devices would need finite output power ${\rm P}{=}-I_{\rm c} \Delta V$ for the successful operation. In addition, the output power can be optimized with respect to the applied voltage bias for the given temperature drop across the impurity. Using the transport integral provided by the Eq.~\eqref{bd11} we cast the maximum power produced by the SU($N$) Kondo correlated nano devices into the form \DK{${\rm P}^{\rm Kondo}_{\rm max}=N/h\times\pi^2A^{\rm Kondo}_0\Delta T^2$, with the factor $A^{\rm Kondo}_0$ characterizing the SU($N$) Kondo impurity,}
\begin{equation}\label{bd11c}
A^{\rm Kondo}_0{=}\frac{1}{36}\frac{\sin^2 \left(\frac{2 \pi  m}{N}\right)\left(\frac{\pi T}{T^{\rm SU(N)}_K}\right)^2}{\sin ^2\left(\frac{\pi  m}{N}\right){+}\frac{1}{3}\left(\frac{\pi T}{T^{\rm SU(N)}_K}\right)^2\frac{N{+}1}{N{-}1}\cos\left(\frac{2 \pi  m}{N}\right)}.
\end{equation} 
While in non-interacting system studied in Ref.~\cite{whitney1} the factor $A_0$ is purely a constant number, the SU($N$) Kondo impurity offers such generalization that also depends on the system properties, $N$, $m$ and $T/T^{\rm SU(N)}_K$. \DK{For the low energy description of problem considered in this work, we set the reference temperature $T{\leq}T^{\rm SU(N)}_K/7$ to fulfill all the assumptions made in Whitney's work.}

\DK{For the quarter filled systems, since the electronic conductance becomes a constant value, the power output is merely fixed by the reference temperature as seen from the Eq.~\eqref{bd11c}. The factor $\mathcal{A}^{\rm Kondo}_0$ characterizing the power output per degeneracy of SU($N$) Kondo effect is plotted (in the unit of $\mathcal{A}_0$) in Fig.~\ref{exceeding} as a function of occupancy $m$ for fixed $N$ (upper panel) and in reverse order (lower panel). As seen from the Fig~\ref{exceeding}, about $50\%$ of quantum upper bound of output power can be generated with SU($N$) Kondo effects. This observation would be further enhanced with the inclusion of nonlinear effects~\cite{dee1}. Though the power output per degeneracy is about the half of the quantum bound, SU($N$) Kondo effect offers the setup with $N{\gg}1$ resulting in a giant output power. Note that, mere increase of $N$ fixing $m$ to the small value would not be useful to increase the power output (see lower panel of Fig.~\ref{exceeding}). This generic feature of strong interplay between the Kondo resonance and filling factor is in striking contrast to the non-interacting system studied in Ref.~\cite{whitney1}. In addition, the optimal value for the coefficient $A^{\rm Kondo}_0$ given in Eq.~\eqref{bd11c} is achieve with the filling factor $m/N{=}1/6$ relevant to the existing proposed realizations~\cite{kita, su12, su6}.

\textit{Conclusions}$-$ We analyzed the maximum power output of a heat engine in SU($N$) Kondo regime based on the local FL theory. The output power per degeneracy with beyond half-filled SU($N$) Kondo effect is observed to be around half of the quantum upper bound. Even though observed power output per degeneracy is already comparable to that observed in experiment, our work pave a way of accessing $\mathcal{K}{\gg}1$ in Whitney's non-interacting formula. In non-interacting systems, at most the spin degeneracy can be utilized to have $\mathcal{K}{=}2$ unlike the system considered in this work, where strong interaction results in the pseudo-spin state with $\mathcal{K}{\gg}1$.  We stress that our predictions can be tested experimentally in various existing setups with CNTs and double QDs. Extending the presented discussion to the multichannel, multistage Kondo paradigm~\cite{dee2, dee3} appears to be a valid avenue for future research.

\textit{Acknowledgements}$-$ We are grateful to Robert Whitney for the illuminating discussions about the Whitney's formula of power production in nano devices during the ICTP summer college on Energy transport and energy conversion in the quantum regime. We are thankful to Liliana Arrachea, Giuliano Benenti, Giulio Casati, Yigal Meir and Yuval Oreg for useful comments. The work of M.K. was  performed  in  part  at  Aspen Center for Physics,
which is supported by National Science  Foundation  grant  PHY-1607611 and
partially supported by a grant from the Simons Foundation.
}

\end{document}